# Scheme for preparation of W state via cavity QED

Zhuo-Liang Cao[*], Ming Yang

Department of Physics, Anhui University, Hefei, 230039, PRChina

**Abstract**

In this paper, we presented a physical scheme to generate the multi-cavity maximally entangled W state via cavity QED. All the operations needed in this scheme are to modulate the interaction time only once.

[*] E-mail address: caoju@mars.ahu.edu.cn

Entanglement is a fundamental concept in quantum mechanics. On one hand, it provides a tool to disprove the local hidden variable theory[1, 2]. On the other hand, the non-locality feature of it brings many applications of entanglement, such as quantum teleportation[3, 4] , quantum dense coding[5, 6] and quantum cryptography[7, 8] etc.

With the development of scientific research, the bipartite entanglement has been studied intensively, and ones can get a rather complete understanding of the nature[2], generation and applications[3, 9-13] of bipartite entangled states. More and more interests have been focused on the research on entanglement of multiparticle[14-17]. In multiparitcle entanglement, there are two different classes of entangled states, the GHZ class state[18] and the W class state[14]. Because they can not be converted into each other by stochastic local operations and classical communications(SLOCC)[14], the two kinds of entangled state do not belongs to the same class. GHZ state is a maximally entangled state in many senses, for instance, it maximally violates the Bell inequalities, the mutual information of measurement outcomes is maximal, it is maximally stable against (white) noise and one can locally obtain from a GHZ state with unit probability an EPR state shared between any two of the three parties. However, when one of the three particles is traced out, the two remaining particles are unentangled. But, for W state, the remaining two particles retain a relative high entanglement degree when one of the three particles is traced out. So the W state is more stable than GHZ state against the particle losses[14].

For the application purpose[3-8], the most important step is to manipulate the entanglement. All the applications are based on the entangled states initially prepared. So the preparation of entanglement is a central task in quantum information theory(QIT). Recently, several schemes for the preparation entangled states have been proposed, using single photon interference[19, 20], cavity QED[10-12, 15], parametric down conversion[21, 22], ion trap[23], and NMR[24, 25]. In particular, the generation schemes for W state have been proposed recently[15 ,26-29]. Guo Yong Xiang et al presented a experimental scheme to prepare a three-photon W state using linear optical elements[28]. Alternatively, Bao Sen Shi et al proposed a scheme for generating W state of paths and W state of polarization photons using multi-port fiber coupler. Using interference between optical beams, Peng Xue et al have proposed a scheme, which can generate W state of atomic ensembles[27]. Guo Ping Guo et al and Guang Can Guo et al have also propose two different schemes to generate W state using cavity QED[15, 26]. In Guo Ping Guo's scheme, the cavity is virtually excited during the preparation process, the requirements on the quality of cavities is greatly loosened and the effective decoherence time of it is greatly prolonged[15]. In Guang Can Guo's proposal, they generated not only three-atom W state but also three-cavity W state[26]. In this paper, we will propose an alternative scheme for the preparation of three cavities W state via cavity QED. In our scheme, we need to modulate the interaction time between atom and cavities only once, which is simpler than that of Guang Can Guo's proposal. As a straight extension, we will discuss the generation of $N$-cavity entangled W state using this scheme.

First, we will discuss the preparation of three-cavity W state. An atom initially prepared in excited state($|e\rangle_a$) and three identical cavities initially prepared in vacuum state are required. The three cavities have been placed in a common plane, and the focus area of the three fields constructs a cylinder area, which is vertical to the common field plane. The schematic diagram is depicted in Fig.1.

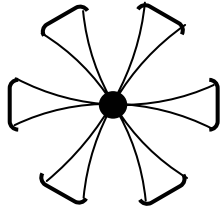

Fig.1. The schematic diagram for generation of W state for three cavities. The black dot denotes the atom.

Then the atom will be sent through the cylinder area from one side(top or bottom) of it. In the focus area, the atom will interact simultaneously with three cavity fields. The interaction can be described by Hamiltonian:

$$\hat{H} = \omega_0 s_z + \sum_{i=1}^{3} \omega_i a_i^+ a_i + \sum_{i=1}^{3} \varepsilon_i \left( a_i s_+ + a_i^+ s_- \right), \tag{1}$$

where $s_z$, $s^+$ and $s^-$ are atomic operators, and $s_z = \frac{1}{2} \left( |e\rangle_a \langle e| - |g\rangle_a \langle g| \right)$, $s^+ = |e\rangle_a \langle g|$, $s^- = |g\rangle_a \langle e|$ with $|e\rangle_a$ and $|g\rangle_a$ being the excited and ground states of the atom. $a_i^+$ and $a_i$ denote the creation and annihilation operators of the $i^{th}$ cavity mode. $\omega_0$ is the transition frequency of the atom and $\omega_i$ is the $i^{th}$ cavity frequency. $\varepsilon_i$ is the coupling constant between the atom and the $i^{th}$ cavity mode. Because the cavities are all identical, if we suppose that the interaction is a resonant one we get: $\varepsilon_1 = \varepsilon_2 = \varepsilon_3 = \varepsilon$ and $\omega_1 = \omega_2 = \omega_3 = \omega_0$.

Before interaction, the state of the total system is:

$$|\Psi(0)\rangle_{123a} = |0\rangle_1 |0\rangle_2 |0\rangle_3 |e\rangle_a. \tag{2}$$

After interaction time $t$, based on Schrödinger equation, the state of the system can be expressed as:

$$|\Psi(t)\rangle_{123a} \rightarrow \cos(\sqrt{3}\varepsilon t) |0\rangle_1 |0\rangle_2 |0\rangle_3 |e\rangle_a$$
$$- i \frac{\sin(\sqrt{3}\varepsilon t)}{\sqrt{3}} \left( |1\rangle_1 |0\rangle_2 |0\rangle_3 + |0\rangle_1 |1\rangle_2 |0\rangle_3 + |0\rangle_1 |0\rangle_2 |1\rangle_3 \right) |g\rangle_a. \tag{3}$$

Then if we select the interaction time $t = \dfrac{\pi}{2\sqrt{3}\varepsilon}$, we can get the maximally entangled W state for three cavities, and the successful probability is 100%, where we have discarded the common phase factor.

Generally, if we can place $N$ cavities in optimal positions, the $N$ cavities W state can also be generated using this scheme. In this case, the Hamiltonian will take a new form:

$$\hat{H} = \omega_0 s_z + \sum_{i=1}^{N} \omega_i a_i^+ a_i + \sum_{i=1}^{N} \varepsilon_i \left( a_i s_+ + a_i^+ s_- \right). \tag{4}$$

The initial state of atom is still $|e\rangle_a$, and the $N$ cavities are also prepared in vacuum state. Then the atom will be sent through the focus area. Similarly, we suppose that the interaction is a resonant one, and all the cavities are identical. Induced by the interaction, the evolution of the system state is:

$$|0\rangle_1 |0\rangle_2 \cdots |0\rangle_N |e\rangle_a$$
$$\xrightarrow{U(t)} \cos(\sqrt{N}\varepsilon t)|0\rangle_1 |0\rangle_2 \cdots |0\rangle_N |e\rangle_a$$
$$-i\frac{\sin(\sqrt{N}\varepsilon t)}{\sqrt{N}}(|1\rangle_1 |0\rangle_2 \cdots |0\rangle_N$$
$$+|0\rangle_1 |1\rangle_2 |0\rangle_3 \cdots |0\rangle_N + \cdots$$
$$+|0\rangle_1 |0\rangle_2 \cdots |0\rangle_{N-1} |1\rangle_N )|g\rangle_a . \quad (5)$$

If the interaction time satisfies: $t = \frac{\pi}{2\sqrt{N}\varepsilon}$, the $N$ cavity will be left in a maximally entangled W state with probability 100%, and the atom is left in ground state.

After generation, we can connect the $N$ cavities with $N$ other cavities distributed in different locations via optical fiber. Then the $N$ distant cavities have been entangled, i.e. $N$-cavity maximally entangled W state has been generated on the $N$ distant cavities. Then the $N$ cavities can be regarded as the nodes on quantum network[30]. Then we can use the W state to realize quantum communication in the network, such as quantum teleportation of unknown atomic state. Teleportation using W state characterize a unique feature that information encoded in the unknown state has been distributed among all the receivers evenly[31].

We proposed a physical scheme in this paper, which can generate three-cavity and multi-cavity maximally entangled W state. Using optical fiber, the generated maximally entangled W state can be distributed among distant users on quantum network[30]. Our proposal is mainly based on cavity QED. Compared with the proposal in Ref[26], our scheme needs fewer operations, which makes it more easily to be realized. But in experiment, there will be a practical problem about how to measure the length of the focus area, which will affect the accuracy of the interaction time. Nevertheless, the simplicity of the scheme makes it more easily to be realized in experiment.

## Acknowledgements

This work is supported by The Project Supported by Anhui Provincial Natural Science Foundation under Grant No: 03042401 and the Natural Science Foundation of the Education Department of Anhui Province under Grant No: 2002kj026, also by the fund of the Core Teacher of Ministry of National Education under Grant No: 200065.

## Reference

1. A. Einstein, B. Podolsky, and N. Rosen, Can quantum-mechanical description of physical reality be considered complete?  **Phys. Rev.** 47, 777 (1935).

2. J. S. Bell, **Physics** (Long Island City, N.Y.) 1, 195 (1965).

3. C. H. Bennett, G. Brassard, C. Crépeau, R. Jozsa, A. Peres, and W. K. Wootter, Teleportation an Unknown Quantum State via Dual Classical and Einstein-Podolsky-Rosen Channels, **Phys. Rev. Lett.** 70, 1895 (1993).

4. D.Bouwmeester, J-W.Pan, et al, Experimental quantum teleportation. **Nature**, 390, 575-579(1997).

5. C. H. Bennett and S. J. Wiesner, Communication via one- and two-particle operators on Einstein-Podolsky-Rosen states, **Phys. Rev. Lett.** 69, 2881(1992).

6. Mattle, K., Weinfurter, H., Kwiat, P. G. & Zeilinger, A. Dense coding in experimental quantum communication. **Phys. Rev. Lett. 76**, 4656-4659 (1996)

7. Ekert, A. Quantum cryptography based on Bell's theorem. **Phys. Rev. Lett.** 67, 661 (1991).

8. C. H. Bennett, G. Brassard, and N. D. Mermin. Quantum cryptography without Bell's theorem, **Phys. Rev. Lett.** 68, 557(1992).

9. P. G. Kwiat Klaus Mattle, Harald Weinfurter, and Anton Zeilinger, New High-Intensity Source of Polarization-Entangled Photon Pairs, **Phys. Rev. Lett**. 75, 4337 (1995).

10. J. M. Raimond, M. Brune, and S. Haroche, Colloquium: Manipulating quantum entanglement with atoms and photons in a cavity, **Reviews of Modern Physics**, 73 565(2001).

11. A. Rauschenbeutel, G. Nogues, S. Osnaghi, P. Bertet, M. Brune, J.M. Raimond, and S. Haroche, **Science** 288, 2024 (2000).

12. E. Hagley, X. Maitre, G. Nogues, C. Wunderlich, M. Brune, J.M. Raimond, and S. Hroche, **Phys. Rev. Lett.** 79, 1 (1997).

13. T. Sleator and H. Weinfurter, Realizable Universal Quantum Logic Gates, **Phys. Rev. Lett. 74**, 4087 (1995).

14. W. Dür, G. Vidal, and JI Cirac, Three qubits can be entangled in two inequivalent ways, **Phys. Rev**. A , 62, 062314 (2000).

15. G. P. Guo , C. F. Li , J. Li, and G. C. Guo, Scheme for the preparation of multiparticle entanglement in cavity QED, **Phys. Rev.** A 65, 042102 (2002).

16. M. Murao, M. B. Plenio, S. Popescu, V. Vedral, and P. L. Knight, Multiparticle entanglement purification protocols, **Phys. Rev.** A 57,  R4075 - R4078 (1998).

17. Adan Cabello, Bell's theorem with and without inequalities for the three-qubit Greenberger-Horne-Zeilinger and W states, **Phys. Rev.** A., 65, (2002) 032108.

18. M. Greenberger, M.A. Horne, A. Shimony, A. Zeilinger, **Am. J. Phys.** 58 1990 1131.

19. Cabrillo, C., Cirac, J. I., G-Fernandez, P. & Zoller, P. Creation of entangled states of distant atoms by interference. **Phys. Rev.** A 59, 1025-1033 (1999).

20. Bose, S., Knight, P. L., Plenio,M. B.& Vedral, V. Proposal for teleportation of an atomic state via cavity decay. **Phys. Rev. Lett.** 83, 5158-5161 (1999).

21. Dik Bouwmeester, Jian-Wei Pan, Matthew Daniell, Harald Weinfurter, and Anton Zeilinger, Observation of Three-Photon Greenberger-Horne-Zeilinger Entanglement, **Phys. Rev. Lett.** 82, 1345–1349 (1999)

22. A. Lamas-Linares et. al. **Nature** 412, 887 (2001)

23. Q. A. Turchette, C. S. Wood, B. E. King, C. J. Myatt, D. Leibfried, W. M. Itano, C. Monroe, and D. J. Wineland, Deterministic Entanglement of Two Trapped Ions, **Phys. Rev. Lett.** 81, 3631–3634 (1998)

24. N. Gershenfeld and I. L. Chuang, **Science** 275, 350(1995)

25. S. L. Braunstein, C. M. Caves, R. Jozsa, N. Linden, S. Popescu, and R. Schack, Separability of Very Noisy Mixed States and Implications for NMR Quantum

Computing **Phys.Rev.Lett.** 83 (1999) 1054-1057

26. Guang-Can Guo and Yong-Sheng Zhang, Scheme for preparation of the W state via cavity quantum electrodynamics, **Phys. Rev.** A 65 054302.

27. Peng Xue, and Guang-Can Guo, Scheme for preparation of mulipartite entanglement of atomic ensembles, **Phys. Rev.** A 67, 034302 (2003)

28. Guo-Yong Xiang, Yong-Sheng Zhang, Jian Li and Guang-Can Guo, Scheme for preparation of the W-state by using linear optical elements, **J. Opt. B: Quantum Semiclass. Opt.** 5 (June 2003) 208-210

29. Bao-Sen Shi, Akihisa Tomita, Schemes for generating W state of paths and W state of polarization photons, **quant-ph**/0208170

30. S. J. van Enk, J. I. Cirac, P. Zoller, H. J. Kimble, H. Mabuchi, **J. Mod. Optics** 44, 1727 (1997).

31. Ye Yeo, Quantum teleportation using three-particle entanglement, **quant-ph**/0302030 (2003).